%% file: paper_LFV.tex
\documentclass[aps,twocolumn,showpacs,preprintnumbers,floatfix,amsmath,amssymb,amsfonts]{revtex4-1}
\pdfoutput=1

\usepackage{mathtools}
\usepackage{atlasphysics}
\usepackage{array}
\usepackage{float}
\usepackage{multirow}
\usepackage{rotating}
\usepackage[caption=false]{subfig} 
\usepackage{cleveref}
\usepackage[hyperindex,breaklinks]{hyperref}

\usepackage{preprintcover}  
\PreprintCoverPaperTitle{Search for the lepton flavor violating decay $Z \rightarrow e\mu$ 
in $pp$ collisions at $\sqrt{s} = 8\TeV$ with the ATLAS detector}  
\PreprintIdNumber{CERN-PH-EP-2014-195}  
\PreprintCoverAbstract{The ATLAS detector at the Large Hadron Collider is used to search for 
the lepton flavor violating process $Z \rightarrow e\mu$ in $pp$ collisions using 
$20.3~\textrm{fb}^{-1}$ of data collected at $\sqrt{s}= 8~\textrm{TeV}$. An enhancement in the $e\mu$ 
invariant mass spectrum is searched for at the $Z$ boson mass. The 
number of $Z$ bosons produced in the data sample is estimated using events of similar 
topology, $Z \rightarrow ee$ and $\mu\mu$, significantly reducing the systematic 
uncertainty in the measurement. There is no evidence of an enhancement at the $Z$ boson mass, 
resulting in an upper limit on the branching fraction, ${\cal B}(Z~\rightarrow~e\mu)~<~7.5 \times 10^{-7}$ at the 95\% 
confidence level.} 
\PreprintJournalName{Physical Review D}

\begin{document}

\title{Search for the lepton flavor violating decay $Z \rightarrow e\mu$ in $pp$ collisions 
at $\sqrt{s} = 8\TeV$ with the ATLAS detector}
\author{The ATLAS Collaboration}

\begin{abstract}
The ATLAS detector at the Large Hadron Collider is used to search for 
the lepton flavor violating process $Z \rightarrow e\mu$ in $pp$ collisions using 
$20.3~\textrm{fb}^{-1}$ of data collected at $\sqrt{s}= 8~\textrm{TeV}$. An enhancement in the $e\mu$ 
invariant mass spectrum is searched for at the $Z$ boson mass. The 
number of $Z$ bosons produced in the data sample is estimated using events of similar 
topology, $Z \rightarrow ee$ and $\mu\mu$, significantly reducing the systematic 
uncertainty in the measurement. There is no evidence of an enhancement at the $Z$ boson mass, 
resulting in an upper limit on the branching fraction, ${\cal B}(Z~\rightarrow~e\mu)~<~7.5 \times 10^{-7}$ at the 95\% 
confidence level.
\end{abstract}

\maketitle

\section{Introduction}
Lepton flavor conservation in the charged lepton sector is a fundamental assumption of the 
Standard Model (SM) but there is no associated symmetry. Thus, searches for lepton flavor violation 
(LFV) processes are good candidates for probing new physics. The observation of neutrino 
oscillations is a clear indication of LFV in the neutral lepton sector; however such an
oscillation mechanism cannot induce observable LFV in the charged lepton sector. All 
searches in the charged lepton sector have produced null results so far~\cite{LFV_review}. Lepton 
flavor violation in the charged lepton sector may have a different origin than LFV induced by neutrino oscillations and the search for this effect provides constraints on
theories beyond the SM (see for example Refs.~\cite{ding,gripaios,hirsch}).

In this paper, a search for the lepton flavor violating decay $Z\rightarrow e\mu$ is 
presented. There are stringent experimental limits on other charged lepton flavor violating 
processes, which can be used to derive an upper limit on the branching fraction for $Z \rightarrow e\mu$
with some theoretical assumptions. For example, the upper limit on $\mu \rightarrow 3e$ yields 
${\cal B}(Z \rightarrow e\mu)  < 10^{-12}$ \cite{Dinh} and on $\mu \rightarrow e\gamma$ 
yields ${\cal B}(Z \rightarrow e\mu) < 10^{-10}$ \cite{davidson}. The experiments at the Large Electron--Positron 
Collider (LEP) searched directly for the decay $Z \rightarrow e\mu$~\cite{akers,delphi,L3,aleph}. The most 
stringent upper limit is ${\cal B}(Z \rightarrow e\mu) < 1.7 \times 10^{-6}$ at the 95\% 
confidence level (CL) using a data sample of $5.0 \times 10^{6}~Z$ 
bosons produced in $e^{+}e^{-}$ collisions at $\sqrt{s} = 88$--$94$ GeV \cite{akers}. The Large Hadron Collider 
(LHC) has already produced many more $Z$ bosons in $pp$ collisions, but with substantially more background. In 
this paper, the $20.3 \pm 0.6~\ifb$~\cite{data} of data collected at $\sqrt{s} = 8~\textrm{TeV}$ by the ATLAS 
experiment corresponds to $7.8 \times 10^{8}~Z$ bosons produced. Despite the larger background at the LHC, a 
more restrictive direct limit on the $Z \rightarrow e\mu$ decay is reported in this paper.

\section{ATLAS Detector}
The ATLAS detector~\cite{atlas} consists of an inner detector (ID) surrounded by a solenoid 
that produces a 2~T magnetic field, electromagnetic and hadronic calorimeters, and a muon 
spectrometer (MS) immersed in a magnetic field produced by a system of toroids. The ID measures the
trajectories of charged particles over the full azimuthal angle and in a pseudorapidity~\footnote{ATLAS 
uses a right-handed coordinate system with its origin at the nominal interaction point (IP) 
in the center of the detector and the $z$-axis along the beam pipe. The $x$-axis points from 
the IP to the center of the LHC ring, and the $y$-axis points upward. Cylindrical coordinates 
$(r,\phi)$ are used in the transverse plane, $\phi$ being the azimuthal angle around the 
beam pipe. The pseudorapidity is defined in terms of the polar angle $\theta$ as 
$\eta=-\ln \textrm{tan}(\theta/2)$. Transverse momentum and energy are defined relative to the beamline as $
\pt=p\sin\theta$ and $E_{\rm{T}}=E\sin\theta$.} range of $|\eta| < 2.5$ using silicon pixel, silicon 
microstrip, and straw-tube transition-radiation tracker (TRT) detectors. Liquid-argon (LAr) 
electromagnetic (EM) sampling calorimeters cover the range $|\eta| < 3.2$ and a 
scintillator-tile calorimeter provides hadronic calorimetry for $|\eta| < 1.7$. 
In the endcaps ($|\eta| > 1.5$), LAr is also used for the hadronic calorimeters, matching the 
outer $|\eta|$ limit of endcap electromagnetic calorimeters. The LAr forward calorimeters 
extend the coverage to $|\eta| < 4.9$ and provide both the electromagnetic and hadronic energy 
measurements. The MS measures the deflection of muons within $|\eta| < 2.7$ using three stations of precision drift 
tubes (with cathode strip chambers in the innermost station for $|\eta| > 2.0$) and provides separate trigger measurements from dedicated chambers in the region $|\eta| < 2.4$.

A three-level trigger system is used to select interesting events to be recorded for 
subsequent offline analysis~\cite{trigger}. For this analysis, the candidate events of interest are 
required to satisfy either a single electron or a single muon trigger that have 
transverse momentum ($p_{\textrm{T}}$) thresholds of $24\GeV$. 

\section{Analysis Strategy}
The event selection requires two high-\pt\, isolated, 
oppositely charged leptons of different flavor: $e^{\pm}\mu^{\mp}$. Events are required 
to contain little jet energy (i.e. small ${p_\textrm{T}}_\textrm{max}^{\textrm{jet}}$, 
the maximum transverse momentum of any jet in an event) and small missing transverse momentum 
(with magnitude $E_{\textrm{T}}^{\textrm{miss}}$). The former eliminates background processes such as $t\overline t \rightarrow e\mu \nu\overline \nu b\overline b$ while the latter rejects $WW \rightarrow e\mu\nu \overline \nu$. These ${p_\textrm{T}}_\textrm{max}^{\textrm{jet}}$ and $E_{\textrm{T}}^{\textrm{miss}}$ requirements are chosen to maximize the Monte Carlo (MC) simulated signal efficiency divided by the square root of the number of candidate background events in the data. Further details of this procedure are given in Sec.~\ref{sec:Event}. After all selection criteria are applied, the dominant background process is $Z \rightarrow \tau\tau \rightarrow e\mu\nu \overline\nu \nu \overline\nu$, which has an $e\mu$ invariant mass ($m_{e\mu}$) spectrum extending into the $Z$ signal region. 

An excess of events above the background expectation is searched for in the $m_{e\mu}$ spectrum at the $Z$ 
boson mass. The number of $Z \rightarrow e\mu$ candidates is estimated by fitting the $m_{e\mu}$ spectrum. The expected signal shape is obtained from MC simulation, while the background is parameterized using a Chebychev polynomial. The branching fraction is obtained from the ratio of the number of observed $Z \rightarrow e\mu$ candidates to the number of observed $Z \rightarrow \ell\ell$ events in the data in the mass range $70 < m_{\ell\ell} < 110\GeV$, where $\ell = e,\mu$. These $Z \rightarrow ee$ and $\mu\mu$ samples are selected with the same selection criteria, resulting in the cancellation of the majority of systematic uncertainties due to electron, muon, and jet reconstruction and modeling. The simulated events are used to cross-check the background level in data and to calculate the selection efficiency for $Z \rightarrow e\mu/ee/\mu\mu$. All selection requirements were fixed before analyzing the data in the $Z$ signal region from $85$ to $95\GeV$. 

\section{Monte Carlo Samples}\label{sec:MC}
Monte Carlo simulated samples normalized to the data integrated luminosity are used to 
determine the major backgrounds pertinent to this analysis as well as to determine the 
optimal $\met$ and ${p_\textrm{T}}_\textrm{max}^{\textrm{jet}}$ 
requirements. All MC samples are produced using the ATLAS detector simulation~\cite{atlas_sim} 
based on GEANT4~\cite{geant}. Signal $Z \rightarrow e\mu$ MC events are produced with 
POWHEG-BOX r1556~\cite{PowhegBox} using the CT10 parton distribution function (PDF)~\cite{CT10} and 
the AU2 set of tunable parameters (tune)~\cite{AU2} along with PYTHIA 8.175~\cite{pythia} for parton showering, 
hadronization and underlying event simulation. To ensure proper normalization of the upper limit to the number 
of~\Zee~and~\Zmm~events, these events are simulated using the same generator as for the signal simulation. In 
practice, the $Z \rightarrow e\mu$ sample is created from a $Z \rightarrow ee$ sample by replacing one of the 
electrons by a muon at the generator level. The $Z \rightarrow \tau\tau$ and $W$ samples are simulated with 
ALPGEN 2.13~\cite{alpgen} interfaced to HERWIG 6.520.2 and PYTHIA 6.426~\cite{pythia6}, respectively, using the 
CTEQ6L1 PDF~\cite{CTEQ6L1} with the AUET2 tune~\cite{AUET2}. The three diboson backgrounds, $q\overline q 
\rightarrow WW$, $gg \rightarrow WW$, and $WZ$, are simulated with the CT10 PDF using \mcatnlo 4.0~
\cite{mc@nlo4manual} with the AUET2 tune, GG2WW~\cite{ggww} with the AUET2 tune, and POWHEG-BOX 
interfaced to PYTHIA 8.165 with the AU2 tune, respectively. The top quark backgrounds, $t\overline t$ 
and single top quark production, are simulated with \mcatnlo 4.0 and AcerMC 3.8~\cite{acer} interfaced to HERWIG 
6.520.2 and PYTHIA 6.426, respectively, for parton showering and fragmentation. An average of 20 additional $pp$ 
collisions per event in the same bunch crossing, known as pileup, are included in each event to match the data. 

\section{Object Selection}\label{sec:Object}
Candidate electrons must have $p_{\textrm{T}}^{e} > 25~\textrm{GeV}$ and, to ensure the shower is 
well contained in the high-granularity region of the EM calorimeter, $|\eta^{e}| < 2.47$~\cite{eID}. The candidate 
must not be in the transition region between the barrel and endcap calorimeters, 
$1.37 < |\eta^{e}| < 1.52$. The impact parameters of the candidate must also be consistent with originating from the 
primary vertex, defined as the reconstructed vertex with the largest sum of track $p_{\textrm{T}}^{2}$, constructed 
from at least three tracks each with $\pt > 400\MeV$. The longitudinal impact parameter, $z_{0}$, measured with 
respect to the primary vertex, of the candidate must satisfy $|z_{0}\sin\theta|< 0.5~\textrm{mm}$ and the transverse 
impact parameter, $d_{0}$, must satisfy $|d_{0}| < 6\sigma_{d_0}$, where $\sigma_{d_0}$ is the uncertainty of the 
impact parameter. The electron candidate must be isolated from other event activity by requiring the sum of the 
transverse momentum of tracks with $p_{\textrm{T}} > 1~\textrm{GeV}$ in a cone of size $\Delta R = \sqrt{(\Delta 
\phi)^{2}+(\Delta \eta)^{2}} = 0.2$ around the candidate to satisfy $\Sigma p_{\textrm{T}}(\Delta R < 0.2)/
p_{\textrm{T}}^{e} < 0.13$. In the calorimeter, the sum of the transverse energy deposits in the calorimeter clusters 
in a cone of size $\Delta R = 0.2$ around the candidate must satisfy 
$\Sigma E_{\rm{T}}(\Delta R < 0.2)/p_{\rm{T}}^{e} < 0.14$. Candidates must also satisfy the ``tight'' identification requirements 
of Ref.~\cite{eID}, which are based on calorimeter shower shape, ID track quality, and the spatial match between the 
shower and the track. 

Muon candidates must have $p_{\textrm{T}}^{\mu} > 25~\textrm{GeV}$ and $|\eta^{\mu}| < 2.5$ to ensure coverage 
by the ID. Muons are required to have a high-quality TRT track segment if they are within the detector 
acceptance of the TRT. To ensure the muon originated from the primary vertex, the distances of closest approach 
to the primary vertex in both $z$ and the transverse plane must satisfy 
$|z_{0}\sin\theta|< 0.5~\textrm{mm}$ and $|d_{0}| < 3\sigma_{d_0}$, respectively. To reject 
secondary muons from hadronic jets, the ID track used in the muon reconstruction 
must be isolated by requiring the sum of the \pt~of the tracks around the muon candidate to satisfy 
$\Sigma p_{\textrm{T}}(\Delta R < 0.2)/p_{\textrm{T}}^{\mu} < 0.15$. In the calorimeter, there should be little activity 
around the muon candidate by requiring the sum of the $E_{\rm{T}}$ around the muon candidate 
to satisfy $\Sigma E_{\rm{T}}(\Delta R < 0.2)/p_{\rm{T}}^{\mu} < 0.3$. Candidates must also satisfy the ``tight'' 
identification requirements of Ref.~\cite{tight} and have their MS track matched to the ID track~\cite{mID}.

Hadronic jets~\cite{jet} are reconstructed using the 
anti-$k_{t}$ algorithm with distance parameter $R = 0.4$~\cite{antikt}. The scalar sum of \pt~of tracks associated with the jet which come from the primary vertex, divided by the scalar sum of \pt~of all tracks associated with the jet, must be greater than $50\%$ for jets with $|\eta| < 2.4$ and $p_{\textrm{T}} < 50~\textrm{GeV}$ to 
remove jets originating from pileup in the central region. The rapidity~\footnote{The rapidity is defined in terms 
of the energy, $E$, and the $z$-component of the momentum along the beam axis, $p_{z}$ as 
$y=\frac{1}{2}\ln \frac{E+p_{z}}{E-p_{z}}$}~of jets must satisfy $|y|<4.4$. 
Finally, only jets with $\pt > 20\GeV$ are considered in the event selection. 

The $E_{\textrm{T}}^{\textrm{miss}}$ is defined as the \pt~imbalance in the detector. It is 
formed from the vector sum of the \pt~of reconstructed high-$p_{\rm{T}}$ objects---electrons, photons, 
jets, $\tau$ leptons, and muons---as well as energy deposits not associated with any reconstructed objects~\cite{met}.

\section{Event Selection}\label{sec:Event}
A $Z$ candidate is constructed from two opposite sign, different flavor leptons ($e$ or $\mu$). 
Electron candidates are vetoed if they are within $\Delta R = 0.1$ of a candidate muon. Jets are 
removed if they are within $\Delta R = 0.3$ of a candidate lepton. Events with more than 
two candidate leptons are vetoed, as are events with an additional electron or muon that 
passed the lepton requirements but is not isolated. 

As stated above, the selection criteria for $E_{\textrm{T}}^{\textrm{miss}}$ and ${p_\textrm{T}}_\textrm{max}
^{\textrm{jet}}$ are chosen to maximize the reconstruction efficiency divided by the square root of the estimated number of background events. The efficiency for selecting $e\mu$ candidates is calculated using MC signal events in the $Z$ signal region, $85~<~m_{e\mu}~<~95\GeV$. The background is determined by 
fitting the $m_{e\mu}$ spectrum in data in the mass range $70 < m_{e\mu} < 110\GeV$, excluding 
the $Z$ signal region, and then interpolating the fitted curve into the $Z$ signal region to estimate the number of background events. The fitting range is chosen so that the $m_{e\mu}$ spectrum can be parameterized with a polynomial. In particular, the lower $m_{e\mu}$ limit is chosen to be above the peak in the $Z \rightarrow \tau\tau \rightarrow e\mu$ mass distribution. The optimum selection criteria are found to be $E_{\textrm{T}}^{\textrm{miss}} <  17~\textrm{GeV}$ and ${p_\textrm{T}}_\textrm{max}^{\textrm{jet}} < 30~\textrm{GeV}$.

Several background functions with a small number of free parameters in the fit were investigated before analyzing (``unblinding'') the events in the $Z$ mass region. This includes Chebychev polynomials of $2^{\textrm{nd}}$ to $4^{\textrm{th}}$-orders, a Landau function, and an exponential function plus a linear term. The $2^{\textrm{nd}}$-order polynomial has an unacceptable $\chi^2$ per degree of freedom, $\chi^2/DOF~=~3.3$. All other functions have $\chi^2/DOF \sim 1$. The $3^{\textrm{rd}}$-order polynomial is chosen as the default background function for simplicity. The systematic error due to the choice of fitting functions is discussed below.

The $E_{\textrm{T}}^{\textrm{miss}}$ and ${p_\textrm{T}}_\textrm{max}^{\textrm{jet}}$ distributions in the data are 
compared with the expectation for a MC simulation of the background and signal in Fig.~\ref{fig:metjetdist2012}. 
Each plot has all kinematic cuts applied with the exception of the cut on the kinematic variable being shown---as 
indicated by the vertical lines and arrows. The signal MC is scaled to the 95\% CL upper limit presented in
Sec.~\ref{sec:Result}. The multijet background in these distributions refers to events where at least two jets are 
misidentified as leptons. The shape and normalization of this background can be estimated from like-sign $e\mu$ 
candidates in the data. The contributions to the same-sign distribution from top quark and $W/Z$ events are estimated using simulation (Sec.~\ref{sec:MC}) and subtracted from the same-sign data. 

The $E_{\textrm{T}}^{\textrm{miss}}$ distribution of $e\mu$ candidate events is shown 
in Fig.~\ref{fig:metjetdist2012}(a). The $E_{\textrm{T}}^{\textrm{miss}}$ requirement removes most of the diboson 
background while retaining the majority of the simulated signal events. The distribution of the 
${p_\textrm{T}}_\textrm{max}^{\textrm{jet}}$ of the candidate events is shown in 
Fig.~\ref{fig:metjetdist2012}(b). The entries in the first bin correspond to events that have no jets passing the jet 
selection requirements described in Sec.~\ref{sec:Object}. The jet veto eliminates most of the 
$t \overline t$ background while maintaining a high reconstruction efficiency for 
$Z \rightarrow e\mu$. The remaining major backgrounds in the $Z$ signal region 
are diboson, multijet, $Z \rightarrow \tau\tau$, and $Z \rightarrow \mu\mu$. For the 
$Z \rightarrow \mu\mu$ background, one of the muons can interact with the detector material 
leading to the muon being misidentified as an electron due to its overlap with a bremsstrahlung photon. The 
$E_{\textrm{T}}^{\textrm{miss}}$ and the ${p_\textrm{T}}_\textrm{max}^{\textrm{jet}}$ distributions of the background 
are well reproduced by the MC simulation. However, in extracting the upper limit on the branching 
fraction for $Z \rightarrow e\mu$, the background is estimated from the data instead of using MC simulation.

\begin{figure}[htbp]
 \centering
 \captionsetup[subfigure]{labelformat=empty}
  \subfloat[]{\includegraphics[width=0.45\textwidth]{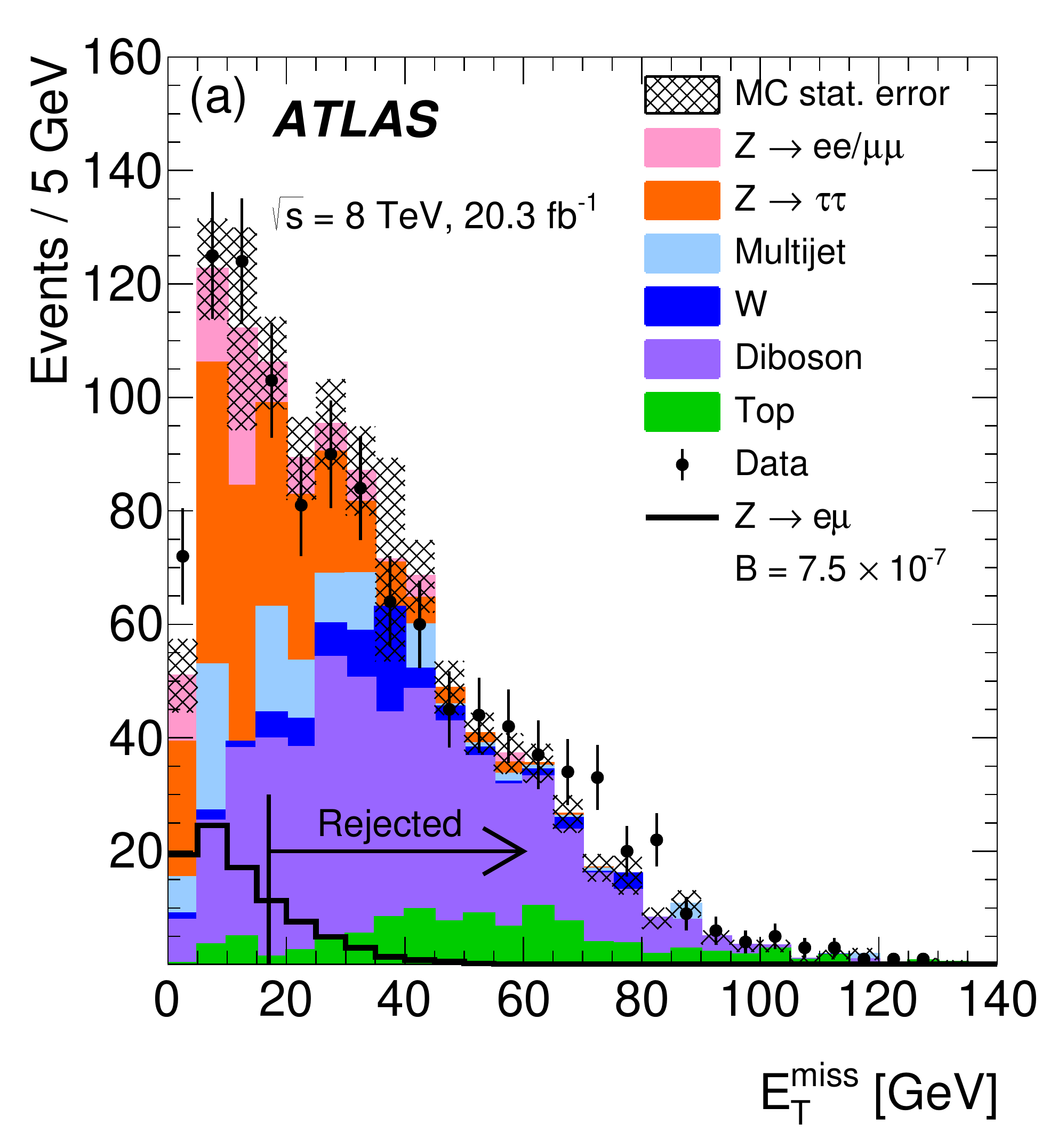}}\\
  \subfloat[]{\includegraphics[width=0.45\textwidth]{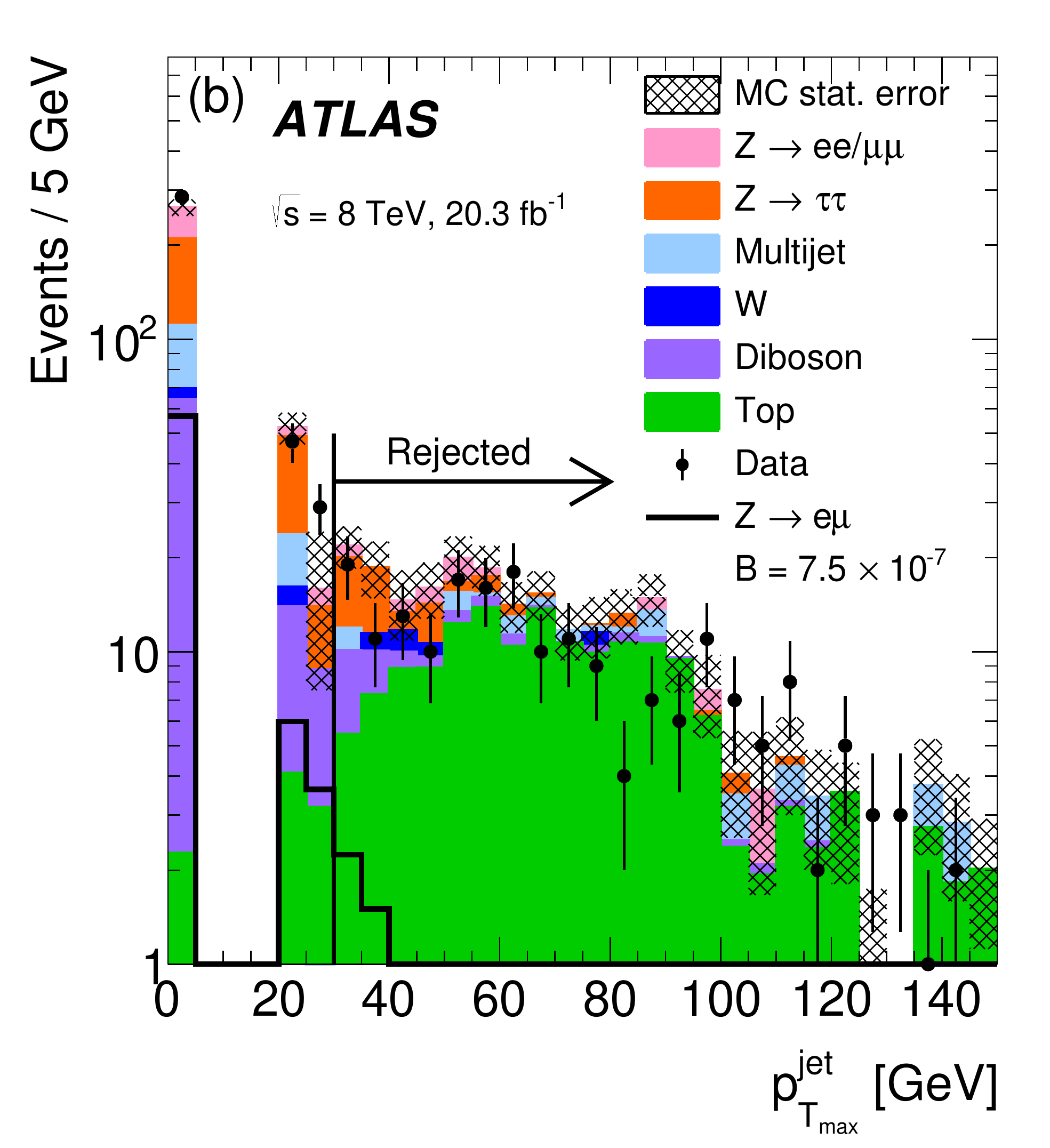} }
 \caption{Distributions of (a) $E_{\textrm{T}}^{\textrm{miss}}$ and (b) ${p_\textrm{T}}_\textrm{max}^{\textrm{jet}}$ 
for $Z\to e\mu$ candidate events with $85~<~m_{e\mu}~<~95~\textrm{GeV}$. The 
expectations for backgrounds from various sources are shown as stacked histograms. Each plot has 
all cuts applied except for the kinematic variable being shown. The vertical lines and arrows indicate 
the $E_{\textrm{T}}^{\textrm{miss}}$ and ${p_\textrm{T}}_\textrm{max}^{\textrm{jet}}$ requirements. 
The hatched bands show the total statistical uncertainty of the backgrounds. The expected distribution of
$Z\to e\mu$ signal events, normalized to the upper limit on the branching fraction (${\cal B}(Z \rightarrow e\mu) = 7.5 \times 10^{-7}$), is indicated by a black line. The entries at zero in the ${p_\textrm{T}}_\textrm{max}^{\textrm{jet}}$ 
distribution correspond to events with no jets that satisfy the jet selection.}
 \label{fig:metjetdist2012}
 \end{figure}

\section{Result}\label{sec:Result}

The $m_{e\mu}$ distribution with the background expectations superimposed is shown in 
Fig.~\ref{fig:massMCdist2012}. The mass spectrum is consistent with the MC background expectation with no 
evidence of an enhancement at the $Z$ mass. The mass spectrum is fit 
as a sum of signal and background contributions as shown in Fig.~\ref{fig:massfit}. The signal shape is a binned 
histogram obtained from the signal MC sample and the absolute normalization is a free parameter in the fit. The 
background is a $3^{\textrm{rd}}$-order Chebychev polynomial function. The fit yields a signal of $4 \pm 35$ events. 

\begin{figure}[htbp]
 \centering
 \includegraphics[width=0.45\textwidth]{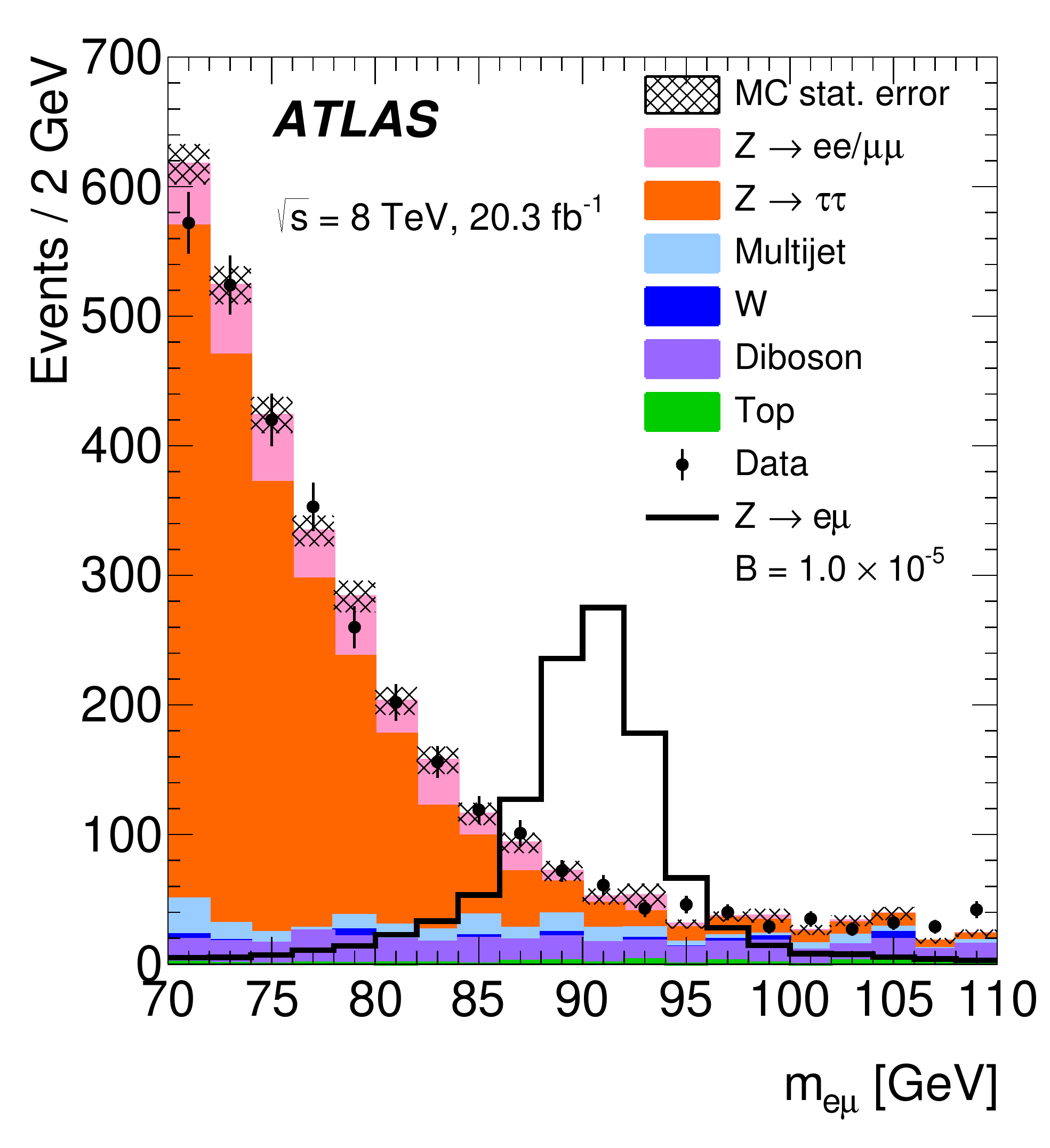}
 \caption{The $e\mu$ invariant mass distribution in data with the background expectations from various 
processes after all cuts applied. The hatched bands show the total statistical uncertainty of backgrounds. The expected distribution of $Z\to e\mu$ signal events, normalized to 13 times the upper limit on the branching fraction ($13 \times {\cal B}(Z \rightarrow e\mu) = 1.0 \times 10^{-5}$), is indicated by a black line.}
 \label{fig:massMCdist2012}
 \end{figure}

\begin{figure}[htbp]
 \centering
 \includegraphics[width=0.45\textwidth]{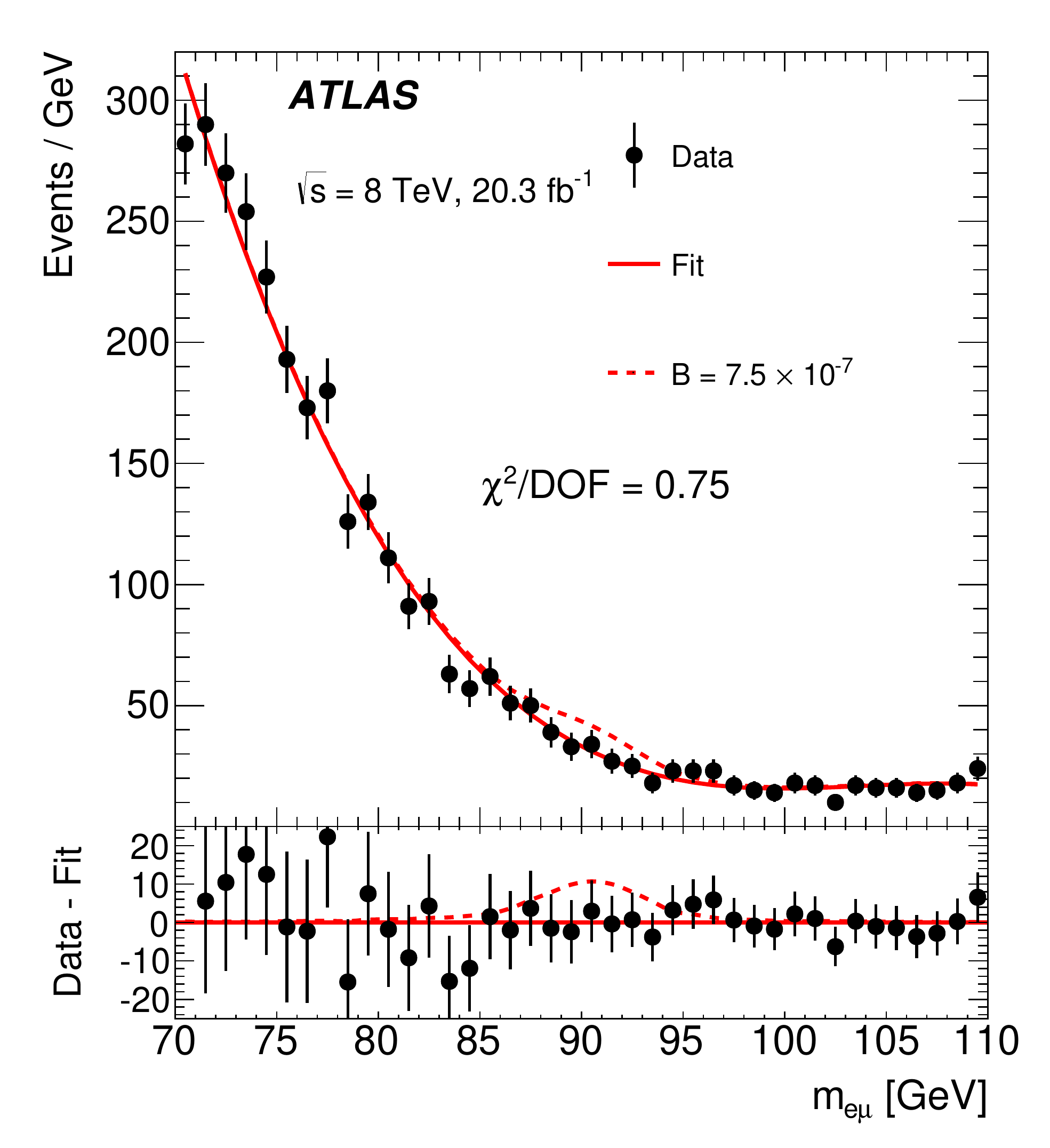}
\caption{The $e\mu$ invariant mass distribution fitted with a signal shape obtained from MC simulation and a $3^{\textrm{rd}}$-order Chebychev polynomial to describe the background (solid). The observed 95\% CL upper limit (dashed) is indicated (${\cal B}(Z \rightarrow e\mu) = 7.5 \times 10^{-7}$). The lower plot shows the data with the background component of the fit subtracted.}
 \label{fig:massfit}
 \end{figure}

The upper limit on ${\cal B}(Z \rightarrow e\mu)$ is given by

\begin{eqnarray}\label{eq:UL}
{\cal B}(Z \rightarrow e\mu) < \frac{N_{95\%}}{\epsilon_{e\mu} N_{Z} },
\end{eqnarray}

\noindent where $N_{95\%}$ is the upper limit on the number of $Z~\rightarrow~e\mu$ candidate events 
at 95\% CL, $\epsilon_{e\mu}$ is the reconstruction efficiency for a $Z \rightarrow e\mu$ event, 
and $N_{Z}$ is an estimate of the total number of $Z$ bosons produced in the data sample. 
This estimate is obtained from the weighted average of two measurements. One is the number of $Z$ bosons 
produced as calculated from the number of $Z \rightarrow ee$ events detected in the data, after correcting for the 
reconstruction efficiency and branching fraction~\cite{PDG}. The other is calculated with the same procedure using 
the $Z \rightarrow \mu\mu$ channel. The numbers of $ee$ and $\mu\mu$ events are estimated by counting the 
candidates with dilepton invariant mass in the region $70~<~m_{\ell\ell}~<~110\GeV$. The reconstruction 
efficiencies are estimated using MC simulation, calibrated with $Z$ candidates using the tag-and-probe 
method~\cite{eID,mID}. The result is summarized in Table~\ref{ta:num2012}. The weight of each measurement is 
given by the total uncertainty, the quadratic sum of the statistical and systematic uncertainties. The systematic 
uncertainties include the uncertainties in the electron and muon reconstruction and trigger efficiencies and the 
absolute scale and resolution of the electron energy and muon \pt~\cite{eIDEscale,mID}. These systematic 
uncertainties are uncorrelated between $ee$ and $\mu\mu$ events. Other systematic uncertainties such as those 
due to imperfect simulation of the $E_{\textrm{T}}^{\textrm{miss}}$ and ${p_\textrm{T}}_\textrm{max}^{\textrm{jet}}$ 
distributions are correlated for the $e\mu$, $ee$, and $\mu\mu$ channels and cancel in the ratio (Eq.~(\ref{eq:UL})), 
although they are major contributors to the systematic uncertainties shown in Table~\ref{ta:num2012} before the 
cancellation. With the cancellation, the systematic uncertainty on ${\cal B}(Z \rightarrow e\mu)$ is 1.2\%, small compared to the overall fitting systematic uncertainty, and is neglected in the final result.

\begin{table}[hbtp]
\vspace{-0.3cm}
\caption{The reconstruction efficiencies for $Z \rightarrow e\mu$, $ee$, and $\mu\mu$ events are shown. 
Also shown are the number of $Z$ bosons produced, $N_{Z}$, as estimated from the number of \Zee~and \Zmm~events, after correcting for the corresponding reconstruction efficiencies and branching fractions, 
as well as the weighted average. The total uncertainties are given.}
\centering
\renewcommand{\arraystretch}{1.2}
\begin{tabular}{c c c}
\hline\hline
$Z$ decay &  Efficiency (\%)  & $N_{Z}$ ($10^{8}$) \\
\hline
$ee$ &  $10.8 \pm 0.3$ &  $7.85 \pm 0.24 $   \\
$\mu\mu$  & $17.8 \pm 0.4$ &  $7.79 \pm 0.17 $ \\
$\langle ee, \mu\mu\rangle$   & &  $7.80 \pm 0.15$     \\
$e\mu$   & $14.2 \pm 0.4$    &            \\
\hline\hline
\end{tabular}
\label{ta:num2012}
\end{table}

A one-sided Profile Likelihood~\cite{roostats} is used as a test statistic to calculate 
an upper limit on the number of signal events using the $CL_{s}$ procedure~\cite{CLs}. The procedure yields an 
observed $95\%$ CL upper limit of 72 events. This is consistent with the expected upper limit of 69 events obtained by generating pseudo-experiments from the observed background spectrum. For the pseudo-experiments, the observed data distribution in the sideband is fitted with a $3^{\textrm{rd}}$-order Chebychev polynomial and the fitted function is then interpolated into the signal region to predict the central value for the number of background events in each bin. The central value of the background events in the background region or interpolated data for the signal region is then fluctuated.

There is a systematic uncertainty due to the choice of fitting function used to 
estimate the background and the associated fitting region (Sec.~\ref{sec:Event}). The upper and lower limits of the fit 
region are varied in the ranges $100$--$120\GeV$ and $70$--$80\GeV$ in $5\GeV$ increments. The background 
parameterization that yields the largest upper limit on the number of signal events (83 events) is used to set an upper limit on the branching fraction at the 95\% confidence level,

\begin{eqnarray}\label{eq:UL2012}
{\cal B}(Z \rightarrow e\mu) < 7.5 \times 10^{-7}.
\end{eqnarray}

\section{Conclusions}
A search for the lepton flavor violating process $Z \rightarrow e\mu$ in $pp$ collisions is 
performed with the ATLAS detector at the LHC. There is no evidence of an enhancement at the $Z$ boson mass in 
the $m_{e\mu}$ spectrum for the dataset with an integrated luminosity of $20.3~\textrm{fb}
^{-1}$ at $\sqrt{s}~=~8\TeV$. Using the $CL_{s}$ method with a one-sided Profile Likelihood as a test statistic, an 
upper limit of 83 signal events at 95\% CL is found. This leads to an upper limit on the branching fraction of ${\cal B}
(Z~\rightarrow~e\mu)~<~7.5 \times 10^{-7}$ at 95\% CL, significantly more restrictive than that from the LEP experiments.

\section*{ACKNOWLEDGMENTS}

We thank CERN for the very successful operation of the LHC, as well as the
support staff from our institutions without whom ATLAS could not be
operated efficiently.

We acknowledge the support of ANPCyT, Argentina; YerPhI, Armenia; ARC,
Australia; BMWF and FWF, Austria; ANAS, Azerbaijan; SSTC, Belarus; CNPq and FAPESP,
Brazil; NSERC, NRC and CFI, Canada; CERN; CONICYT, Chile; CAS, MOST and NSFC,
China; COLCIENCIAS, Colombia; MSMT CR, MPO CR and VSC CR, Czech Republic;
DNRF, DNSRC and Lundbeck Foundation, Denmark; EPLANET, ERC and NSRF, European Union;
IN2P3-CNRS, CEA-DSM/IRFU, France; GNSF, Georgia; BMBF, DFG, HGF, MPG and AvH
Foundation, Germany; GSRT and NSRF, Greece; ISF, MINERVA, GIF, I-CORE and Benoziyo Center,
Israel; INFN, Italy; MEXT and JSPS, Japan; CNRST, Morocco; FOM and NWO,
Netherlands; BRF and RCN, Norway; MNiSW and NCN, Poland; GRICES and FCT, Portugal; MNE/IFA, Romania; MES of Russia and ROSATOM, Russian Federation; JINR; MSTD,
Serbia; MSSR, Slovakia; ARRS and MIZ\v{S}, Slovenia; DST/NRF, South Africa;
MINECO, Spain; SRC and Wallenberg Foundation, Sweden; SER, SNSF and Cantons of
Bern and Geneva, Switzerland; NSC, Taiwan; TAEK, Turkey; STFC, the Royal
Society and Leverhulme Trust, United Kingdom; DOE and NSF, United States of
America.

The crucial computing support from all WLCG partners is acknowledged
gratefully, in particular from CERN and the ATLAS Tier-1 facilities at
TRIUMF (Canada), NDGF (Denmark, Norway, Sweden), CC-IN2P3 (France),
KIT/GridKA (Germany), INFN-CNAF (Italy), NL-T1 (Netherlands), PIC (Spain),
ASGC (Taiwan), RAL (UK) and BNL (USA) and in the Tier-2 facilities
worldwide.

\clearpage
\bibliographystyle{h-physrev}

\onecolumngrid
\clearpage
\input{atlas_authlist}

\end{document}

%% file: atlas_authlist.tex
\begin{flushleft}
{\Large The ATLAS Collaboration}

\bigskip

G.~Aad$^{\rm 85}$,
B.~Abbott$^{\rm 113}$,
J.~Abdallah$^{\rm 153}$,
S.~Abdel~Khalek$^{\rm 117}$,
O.~Abdinov$^{\rm 11}$,
R.~Aben$^{\rm 107}$,
B.~Abi$^{\rm 114}$,
M.~Abolins$^{\rm 90}$,
O.S.~AbouZeid$^{\rm 160}$,
H.~Abramowicz$^{\rm 155}$
% ...
\bigskip
\\
% ...
$^{11}$ Institute of Physics, Azerbaijan Academy of Sciences, Baku, Azerbaijan\\
% ...
$^{85}$ CPPM, Aix-Marseille Universit{\'e} and CNRS/IN2P3, Marseille, France\\
% ...
$^{90}$ Department of Physics and Astronomy, Michigan State University, East Lansing MI, United States of America\\
% ...
$^{107}$ Nikhef National Institute for Subatomic Physics and University of Amsterdam, Amsterdam, Netherlands\\
% ...
$^{113}$ Homer L. Dodge Department of Physics and Astronomy, University of Oklahoma, Norman OK, United States of America\\
$^{114}$ Department of Physics, Oklahoma State University, Stillwater OK, United States of America\\
% ...
$^{117}$ LAL, Universit{\'e} Paris-Sud and CNRS/IN2P3, Orsay, France\\
% ...
$^{153}$ Institute of Physics, Academia Sinica, Taipei, Taiwan\\
% ...
$^{155}$ Raymond and Beverly Sackler School of Physics and Astronomy, Tel Aviv University, Tel Aviv, Israel\\
% ...
$^{160}$ Department of Physics, University of Toronto, Toronto ON, Canada\\
% ...
\end{flushleft}
